\documentstyle[12pt]{article}

\begin{document}
\input{epsf.sty}
\baselineskip 15.2 pt

\title{Teleportation: Dream or Reality?}

\author{Lev Vaidman}

\maketitle

\begin{center}
{\small \em School of Physics and Astronomy \\
Raymond and Beverly Sackler Faculty of Exact Sciences \\
Tel Aviv University, Tel-Aviv 69978, Israel. \\}
\end{center}

\begin{abstract}
  Since its discovery in 1993, we witness an intensive theoretical and
  experimental effort centered on teleportation. Very recently it was
  claimed in the press that ``quantum teleportation has been achieved in the
  laboratory'' (T. Sudbery, {\it Nature} {\bf 390}, 551). Here, I
  briefly review this research focusing on the connection to {\it
    nonlocal measurements}, and question Sudbery's statement. A
  philosophical inquiry about the paradoxical meaning of teleportation
  in the framework of the many-worlds interpretation is added.
\end{abstract}

\section{The meaning of the Word ``Teleportation''}
\label{Intro}

Let me start with a citation from 
the Oxford English Dictionary \cite{OX}:

\begin{quotation}
\noindent  {\bf teleportation}. {\it Psychics} and {\it Science
    Fiction}.
 The conveyance
  of persons (esp. of oneself) or things by psychic power; also in
  futuristic description, apparently instantaneous transportation of
  persons, etc., across space by advanced technological
means. 
\end{quotation}
Recently, the word ``teleportation'' has appeared outside of the realm 
of mystical and science fiction literature: in science journals. 
Bennett, Brassard, Crepeau, Jozsa, Peres, and Wootters (BBCJPW)
\cite{BBC} proposed a gedanken experiment they termed  ``quantum
teleportation''.

Classically, to move a person is to move all the particles it is made
of. However, in quantum theory particles themselves do not represent a
person: all objects are made of the same elementary particles. An
electron in my body is identical to an electron in the paper
of the page you are reading now. An object is characterized by the
{\it quantum state} of the particles it is made of. Thus,
reconstructing the quantum state of these particles on other particles of
the same kind at a remote location {\em is} ``transportation'' of the
object.

The quantum state of the object to be transported is supposed to be
unknown. Indeed,
usually we do not know and cannot find out what the quantum state
of an object is.  Moreover, frequently an object is not in a pure quantum
state, its particles may be correlated to other systems. In such
cases the essence of the object is these correlations. In order to
transport such correlations (even if they are known), without access to
the systems which are in correlation with our system, a method for
teleportation of an unknown quantum state is necessary.

Quantum teleportation \cite{BBC} transfers the quantum state of a
system and its correlations to another system. Moreover, the
procedure corresponds to the modern meaning of teleportation: an
object is disintegrated in one place and a perfect replica appears at
another site.  The object or
its complete description is never located between the two sites during
the transportation. Note that ``disintegration'' of the quantum state is
a necessary requirement due to the no-cloning theorem.

 The teleportation procedure, apart from preparing
in advance the quantum channels, requires telegraphing surprisingly small
amounts of information between the two sites. This stage prevents
``instantaneous'' transportation. Indeed, because of special
relativity, we  cannot hope to achieve superluminal teleportation: 
 objects  carry  signals.

Due to the arguments presented above, I find the BBCJPW procedure to
be very close to the concept of ``teleportation'' as it is used in the
science-fiction literature.  However, the name teleportation is less
justified for the recent implementations of this idea in the laboratory,
as well as for some other proposals for experiments.  For me, an
experiment deserves the name ``teleportation'' if I can give to Alice
(the sender) a system whose quantum state in unknown to her and that
she can, without moving this system and without moving any other
system which can carry the quantum state of the system, transport this
state to Bob (the receiver) which is located at a remote location.  In
the next section I shall discuss, in the light of my definition, the
usage of the word ``teleportation''.

\section{Recent Experiments and Proposals for Experiments Termed
  ``Teleportation''}
\label{examples}

What I discuss in this section is essentially a
semantic issue, but I feel that its clarification is important. I find
the original teleportation paper to be one of the most important
results in the field in the last ten years, and I think that it should
be clearly distinguished from other interesting but less profound achievements.

Recently I heard the word ``teleportation'' in the context of NMR-type
quantum computation experiments \cite{NMR-tele}. Using certain pulses,
a spin state of a nucleus in a large molecule is transported to
another nucleus in the same molecule. The main deficiency of this
experiment as teleportation is that it does not allow to transport an
{\em unknown} quantum state. Indeed, in the NMR experiments a
macroscopic number of molecules have to be in a particular quantum
state. If Alice receives a single quantum object in an unknown quantum
state, she cannot duplicate it and in that manner prepare many copies
in many molecules, due to the no-cloning theorem.

An apparent  weakness of the NMR experiment is that the internal coupling
which plays the role of the channel for classical information required
for teleportation {\em can}, in principle, carry the quantum state.
However, due to the strong interaction with the environment, the quantum
state transmitted through such a channel is effectively measured by
the environment. Only the eigenstates corresponding to the classical
outcomes are stable under this interaction and, therefore, there is 
good reason to consider this channel to be classical.

Another place in which I encountered the word ``teleportation'' is the
work on optical simulation of quantum computation \cite{cerf}. It
includes a proposal for implementation of the idea to view
``teleportation'' as a particular quantum computation circuit
\cite{Brass,BBC}.  The problem in the optical experiment is that
instead of the classical channel which is supposed to transmit two
bits of information, real photons are moving from Alice to Bob and
these photons {\em can} transmit the whole quantum state of the
polarization degree of freedom of the photon. This is exactly the
apparent weakness of the NMR-teleportation experiments mentioned
above, but in the present case the environment does not make the
quantum channel to be effectively classical. Note that in the original
proposal the quantum channel is explicitly replaced by a classical one
to make the proposal akin to teleportation in the BBCJPW sense.  It is
the optical simulation of this proposal which is something less than
teleportation. It seems that the authors \cite{cerf} were aware of
this problem when they added a footnote: ``The term teleportation is
used in the literature to refer to the transfer of the state of a
qubit to another''. I find this meaning to be too general. Many
processes corresponding to this definition were proposed (and even
implemented in laboratories) long before the teleportation paper has
appeared.

Next, let me discuss ``teleportation'' in the Rome experiment
\cite{Rome}. As I will explain in the next section, the main obstacle
for successful reliable teleportation is the experimental difficulty
to make one quantum object interact with another. In optical
experiments quantum objects, photons, interact with classical objects
such as beam splitters, detectors, etc.  Popescu \cite{Pop} proposed a
very elegant solution: two degrees of freedom of a {\em single} photon
do interact effectively one with the other.  This idea was
successfully implemented in the Rome experiment in which the
polarization state of the photon was transported to another photon.
However, the weakness of this experiment is that the quantum state to
be teleported has to be the state of (the second degree of freedom of)
one of the members of the EPR pair which constitute the quantum
channel of the teleportation experiment.  Therefore, this method
cannot be used for teleportation of an unknown quantum state of an
external system.  The authors \cite{Rome} view this experiment as
``teleportation'' because after the preparation Alice cannot find out
the quantum state, which, nevertheless, is transported (always and
with high fidelity) to Bob.

Finally, let me discuss the Innsbruck teleportation experiment
\cite{Inn,swap}.  Although the word ``teleportation'' appears in the title of
the first Letter \cite{Inn},  the second experiment
\cite{swap}  is a much better
demonstration of teleportation. I believe that the Innsbruck
experiment deserves the name teleportation. It showed for the first
time that  an
unknown state of an external photon can be teleported. It is not a reliable
teleportation: the experiment has a theoretical success rate of 25\%
only, and the employed methods cannot, in principle, lead to reliable
teleportation. For a system consisting of $N$ qubits the probability
of successful teleportation is exponentially small.

Recently, Braunstein and Kimble \cite{BK-N} pointed out a weak point
of the Innsbruck experiment. In the current version of the experiment
one might know that teleportation has been successful only after the time Bob
detects (and, therefore, destructs) the photon with the teleported state.
Thus, the name given by Braunstein and Kimble for the Innsbruck
experiment: ``a posteriori teleportation'' appears to be appropriate.
However, as mentioned in the reply \cite{reply} and in the comment
itself, it is feasible to solve this problem  by a  modification of the
experiment and therefore it is not a conceptual difficulty.

Another possible improvement of the demonstration of teleportation in
the Innsbruck experiment is using single input photons. In the current
version of the experiment, the polarizer which controls the input
quantum state is stationary, and, therefore, many photons are created
in the same state. Thus, this state can hardly be considered an
``unknown'' quantum state. Low intensity of the input beam and frequent
changing of the angle of the polarizer is a simple and effective 
solution of the problem. An ideal solution is using a ``single-photon
gun'' \cite{SFG} which creates  single-photon states.

Apart from the impossibility of performing a measurement of the
nondegenerate Bell operator, there is another problem for achieving
reliable teleportation of an unknown state of a single photon. Today,
there is no source which creates a single EPR pair at will, something
frequently called an ``event-ready'' source. The second Innsbruck
experiment \cite{swap} is the best achievement in this direction:
entanglement swapping may be viewed as creation of an entangled
pair at the moment of the coincidence detection of the two photons coming
from the beam splitter.  What is missing is a ``sophisticated detection
procedure'' \cite{swap} which rules out the creation of two pairs in a
single crystal.

\section{Bell-Operator Measurement and Teleportation}
\label{proof}

The original BBCJPW teleportation procedure consists of three main
stages:
(i)~Preparation of an EPR pair,
(ii)  Bell-operator measurement performed on the ``input'' particle and one
particle of the EPR pair,
(iii) Transmission of the outcome of the Bell measurement and
appropriate unitary operation on the second particle of the EPR pair
(the ``output'' particle).
Completing (i)-(iii) ensures transportation of the pure state of the
input particle to  the output particle. It also ensures
transportation of correlations: if the input particle were correlated
to other systems, then the output particle ends up correlated to these
systems in the same way.

The main difficulty in this procedure is performing the Bell
measurement. Recently it has been proved \cite{VY,FIN} that without
``quantum-quantum'' interaction one cannot perform measurement of the
nondegenerate Bell operator which is required for reliable
teleportation. Using only ``quantum-classical'' interactions one  can
perform a measurement of a degenerate Bell operator
\cite{Harold,BrMa}, thus allowing a teleportation which succeeds
 sometimes.

The size limitations of this paper allow  only to outline  the proof \cite{VY}.  In order to prove that it is impossible
to perform complete (nondegenerate) Bell-operator measurements without
using interactions between quantum systems, I  assume that any
unitary transformation of single-particle states and any local
single-particle measurement are allowed.
There are four distinct (orthogonal)
single-particle states  involved in the definition of the
Bell states:  two channels, and a two-level system which enters
into each channel.  We  name the channels left (L) and right (R),
corresponding to the way the Bell states  are written:
\begin{eqnarray}
\label{Bell}
\nonumber |\Psi_{\pm}\rangle  &=&{1\over\sqrt2}
(|{\uparrow}\rangle_L|{\downarrow}\rangle_R {\pm}
|{\downarrow}\rangle_L|{\uparrow}\rangle_R),
\\
|\Phi_ {\pm}\rangle &=&{1\over \sqrt2}(|{\uparrow}\rangle_L|{\uparrow}\rangle_R  {\pm}
|{\downarrow}\rangle_L|{\downarrow}\rangle_R).
\end{eqnarray}

\noindent
The measurement
procedure can be divided into two stages: the unitary linear evolution,
and local detection.
The general form of the
unitary linear evolution of the four single-particle
states can be written in the following manner:
\begin{equation}
\label{basic}
 |{\uparrow}\rangle_L \rightarrow \sum a_i |i\rangle, 
~~ |{\downarrow}\rangle_L \rightarrow \sum b_i |i\rangle,~~
|{\uparrow}\rangle_R \rightarrow \sum c_i |i\rangle,
~~ |{\downarrow}\rangle_R \rightarrow \sum d_i |i\rangle,
\end{equation}
where $\{|i\rangle\}$ is a set of orthogonal single-particle local
states.  The ``linearity'' implies that the evolution of the particle in
one channel is independent on the state of the particle in another
channel and, therefore, Eq. (\ref{basic}) is enough to define the
evolution of the Bell states:
\begin{eqnarray}
\label{unit}
\nonumber 
 |\Psi_-\rangle
    &\rightarrow &\sum_{i,j} \alpha_{ij} |i\rangle |j\rangle,
~~
 |\Psi_+\rangle 
\rightarrow \sum_{i,j} \beta_{ij}  |i\rangle |j\rangle,
\\
|\Phi_-\rangle 
&\rightarrow &\sum_{i,j} \gamma_{ij} |i\rangle |j\rangle,
~~
 |\Phi_+\rangle 
\rightarrow \sum_{i,j} \delta_{ij} |i\rangle |j\rangle.
\end{eqnarray}
Proper symmetrization is  required for
identical particles.

I assume that there are only local detectors and, therefore, only 
product states $ |i\rangle |j\rangle$ (and not their superpositions)
can be detected.  Measurability of the non-degenerate Bell operator
means that there is at least one nonzero coefficient of every kind
$\alpha_{ij}, \beta_{ij}, \gamma_{ij}, \delta_{ij}$ and if, for a
certain $i,j$, it is not zero, then all others are zero.
This observation leads to numerous equations which, after some tedious
algebra, yield the desired proof.

A somewhat different approach was taken in proof \cite{FIN}. This
proof considers only photons, but it proves the impossibility of
non-degenerate Bell-measurements for a more general case in which
measurements in two stages are allowed.  The procedure in which the
choice of the measurements in the second stage depends on the results
of the measurements in the first stage is an indirect quantum-quantum
interaction: the state of one quantum system influences the result of
the first measurement and the action on the second quantum system
depends on this result.

If we allow direct quantum-quantum interactions, we can achieve
reliable (theoretically 100\% efficient) teleportation. In this case,
we can perform a measurement of the non-degenerate Bell operator.
Indeed, a quantum-quantum interaction such as a conditional spin-flip
transforms the Bell states into product states which then can be
measured using single-particle measuring devices.

An alternative method of teleportation \cite{tele-V} is based on  {\em
  nonlocal} measurements \cite{AAV86} ``crossed'' in space-time.  In order to
teleport a quantum state from particle 1 to particle 2 and, at the
same time, the quantum state of particle 2 to particle 1, the following
(nonlocal in space-time) variables  should be measured (see Fig. 1):
\begin{equation}
\label{swap}
{\cal Z} \equiv  \Bigl({\sigma_1}_z(t_1) + {\sigma_2}_z(t_2)\Bigr){\rm
  mod~4}, ~~
{\cal X} \equiv \Bigl({\sigma_1}_x(t_2) + {\sigma_2}_x(t_1)\Bigr){\rm
mod~4}.
\end{equation}
For any set of outcomes of the nonlocal measurements (\ref{swap}) the
spin state is teleported; in some cases the state is rotated by $\pi$
around one of the axes, but the resulting rotation can be inferred
from the nonlocal measurements.  In order to perform nonlocal
measurements (\ref{swap}), correlated pairs of auxiliary particles
located at the sites of particle 1 and 2 are required.  For completing
the whole procedure we need two singlets instead of the one required in
the original teleportation procedure. The reason for requiring more
resources is that two-way (rather than one-way) teleportation is
achieved.

\begin{center} \leavevmode \epsfbox{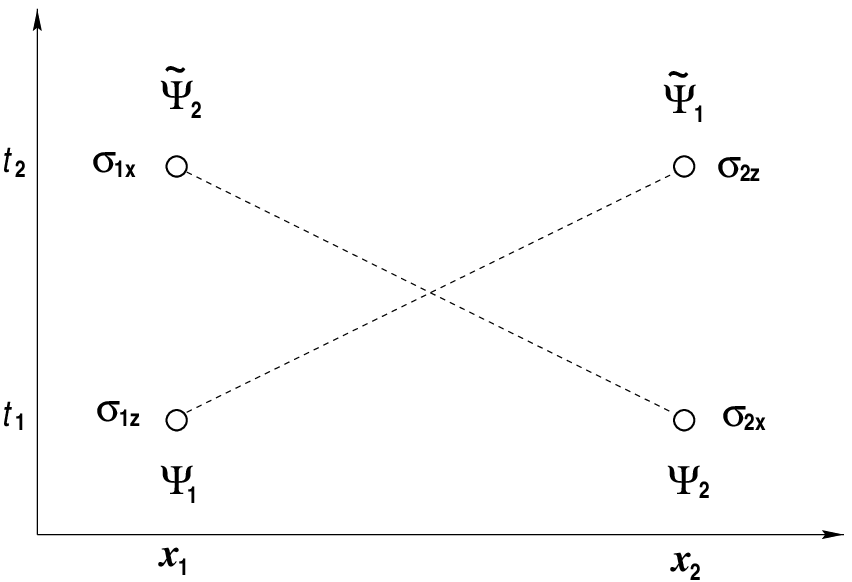} \end{center}

\noindent 
{\small {\bf Fig. 1. ~ Space-time diagram of ``crossed'' nonlocal
  measurements which result in two-way teleportation.}~ 
 Space-time locations of local couplings are
  shown.  When the nonlocal measurements (\ref{swap}) are completed,
  the states of the two particles are interchanged up to local $\pi$
  rotations, $\tilde \Psi_i$ signifies ``rotated'' $\Psi_i$.}


\section{Towards experimental realization of reliable  teleportation }
\label{tele-exp}

Due to the lack of an effective photon-photon interaction, the
currently available methods do not allow reliable teleportation of
the photon polarization state.  It seems that the most promising
candidates for teleportation experiments which might have 100\%
success rate are proposals which involve atoms and electro-magnetic
cavities. First suggestions for such experiments \cite{Slea,David,Cirac} were made shortly
after publication of the original teleportation paper
 and numerous modifications appeared since.
The  implementation of these proposals seems to be
feasible because of the existence of  the ``quantum-quantum'' interaction
between the system carrying the quantum state and a system forming the
EPR pair.  A dispersive interaction (DI) of a Rydberg atom
passing through a properly tuned micro-wave cavity leads to a
conditional phase flip depending on the presence of a photon in the
cavity. A resonant interaction (RI-$\pi$) between the Rydberg atom and
the cavity allows swapping of quantum states of the atom and the
cavity. Thus, manipulation of the quantum state of the cavity can be
achieved via manipulation of the state of the Rydberg atom. The atom's
state is transformed by sending it through an appropriately tuned
microwave zone. Moreover, the direct analog of conditional spin-flip
the interaction can be achieved through the Raman atom-cavity interaction
\cite{ZG97}. No teleportation experiment has been performed as of yet
using these methods, but it seems that the technology is not too far
from this goal.  Recent experiments on atom-cavity interactions
\cite{Har} teach us about the progress
 in this direction.

Until further progress in technology is achieved, it is not easy to
predict which proposal will be implemented first.  Assuming that
resonant atom-cavity interactions can be performed with very good
precision and that a dispersive interaction is available with a
reasonable precision, it seems that the following is the simplest
proposal, see Fig.~2.  The quantum channel consists of a cavity and a
Rydberg atom in a correlated state.  A particular resonant
interaction, RI-$\pi / 2$, of an excited atom passing through an empty
cavity,
 \begin{equation}
   \label{resona}
\nonumber
|e\rangle |0\rangle \rightarrow  {1\over \sqrt2}(|g\rangle |1\rangle +
|e\rangle |0\rangle)  ,
 \end{equation}
prepares this quantum channel (Fig. 2a).  The quantum state to be
teleported is the state of another Rydberg atom. The Bell measurement
is then performed on this atom and the cavity. To this end, the atom
passes through the cavity interacting dispersively (Fig. 2b), induces
the conditional phase flip,
\begin{equation}
\label{pha-fli-ca}
|e\rangle|0\rangle \rightarrow 
|e\rangle|0\rangle,
~~~
|e\rangle|1\rangle \rightarrow -
|e\rangle|1\rangle,
~~~
|g\rangle|0\rangle \rightarrow 
|g\rangle|0\rangle,
~~~
|g\rangle|1\rangle \rightarrow 
|g\rangle|1\rangle,
\end{equation}
which  disentangles the following Bell states:
\begin{eqnarray}
\label{Bell-ca}
\nonumber |\Psi_{\pm}\rangle  &=& {1\over 2}\Bigl(|e\rangle (|0\rangle
-|1\rangle)  \pm |g\rangle (| 1\rangle +|0\rangle)\Bigr),
\\
 |\Phi_{\pm}\rangle  &=& {1\over 2}\Bigl(|e\rangle (|0\rangle
+|1\rangle)  \pm |g\rangle (| 1\rangle -|0\rangle)\Bigr).
\end{eqnarray}

\begin{center} \leavevmode \epsfbox{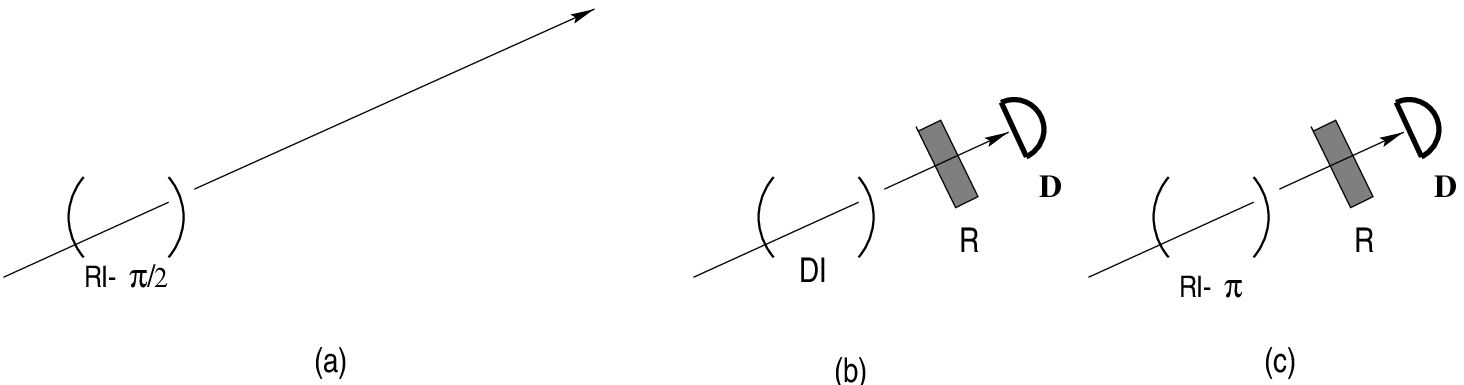} \end{center}

\noindent 
{\small {\bf Fig. 2.~  Single-cavity teleportation of a quantum state of an
  atom.}\\ 
  (a) Preparation of the quantum channel. An atom undergoes resonant interaction
  RI-$\pi$/2 with the cavity and moves to a remote site. 
 \\ (b) The atom, carrying the quantum state to be teleported, interacts
 with the cavity dispersively and its state is measured.
  \\
  (c) The state of the cavity is measured using an auxiliary atom.}
\vskip .4cm

\noindent
The Bell states (\ref{Bell-ca})  have the form of Eq. \ref{Bell} when the first
$|{\uparrow}\rangle$ in the product  is
identified with $|e\rangle$, the second $|{\uparrow}\rangle$,  with
 $(1/\sqrt2)(|0\rangle +
|1\rangle)$, etc.  Measurement of the atom state and the cavity state
 completes the Bell measurement procedure.
In order to make the measurement of the cavity state we perform another
  resonant interaction, RI-$\pi$,  between the cavity and an auxiliary
  atom prepared initially in the ground
state (Fig. 2c),   
 \begin{equation}
   \label{resona1}
|g\rangle |1\rangle \rightarrow  |e\rangle |0\rangle, ~~~
 |g\rangle |0\rangle \rightarrow  |g\rangle |0\rangle. 
 \end{equation}
This interaction transfers the quantum
state of the cavity to this atom. The final measurements on the atoms
 distinguish between the states
  $(1/\sqrt2)(|g\rangle +
|e\rangle)$ and $(1/\sqrt2)(|g\rangle - 
|e\rangle)$.
Since detectors can distinguish between $|g\rangle$ and $|e\rangle$,
the states of the atoms are rotated while  passing through the appropriate
microwave zones before detection. When the Bell measurement is
completed, the quantum state is teleported up to the known local
transformation determined by the results of the Bell measurement. (This
final local transformation is not shown in Fig. 2.)

One relatively simple method for ``two-way'' teleportation of atomic
states is a direct implementation of the crossed nonlocal measurement
scheme presented in the previous section. This method is described
in Ref. \cite{VY}.

One difficulty with the teleportation of atomic states is that usually
experiments are performed with atomic {\em beams} and not with
individual atoms. Such experiments might be good for demonstration and
studying experimental difficulties of teleportation, but they cannot
be considered as implementation of the original wisdom of
teleportation or used for cryptographic purposes. In fact, optical
experiments have this difficulty too, unless ``single-photon guns''
will be used. Both for atomic and for optical experiments this
difficulty does not seem to be unsolvable, but it certainly brings
attention to experiments with trapped ions \cite{ion}. There are many
similarities between available manipulations with atoms and with ions,
so the methods discussed above might be implemented for ion systems
too.

Note also another  recent proposal for teleportation using
quantum-quantum interaction \cite{chiral}. It is  based on rotation of the photon
polarization due to presence of a single   chiral molecule in an
optical cavity. I am, however, skeptical about the  feasibility of such
experiment due to difficulties in tuning the interferometer  
in which photons undergo multiple reflections in the cavity; the
number of reflections has  to be very large due to   weakness of the
interaction between the molecule and the photon.

\section{Teleportation of continuous variables}
\label{tele-cont}

In the framework of nonlocal measurements there is a natural way of
extending the teleportation scheme to systems with continuous
variables \cite{tele-V}.  Consider two similar systems located far
away from each other and described by continuous variables $q_1$ and 
$q_2$ with corresponding conjugate momenta $p_1$ and $p_2$. In order
to teleport the quantum state of the first particle    $\Psi_1(q_1)$ to
the second particle (and the state of the second particle $\Psi_2(q_2)$ to the first) we perform the following
``crossed'' nonlocal measurements (see Fig. 3), obtaining the
outcomes $a$ and $b$:

 \begin{equation}
\label{cross-conti} 
q_1(t_1) - q_2(t_2) = a,~~~
 p_1(t_2) - p_2(t_1) = b. 
\end{equation}
In Ref. \cite{tele-V} it is shown that these nonlocal ``crossed''
measurements ``swap'' the quantum states of the two particles up to
the known shifts in $q$ and $p$. Indeed, the  states of the particles
after completion of 
the measurements (\ref{cross-conti}) are
\begin{equation}
\label{psi-shif}
\Psi_{f}(q_1)= e^{ibq_1} \Psi_2(q_1+a),~~~
 \Psi_{f}(q_2)= e^{-ibq_2} \Psi_1(q_2-a) . 
\end{equation}
\vskip .4cm

\begin{center} \leavevmode \epsfbox{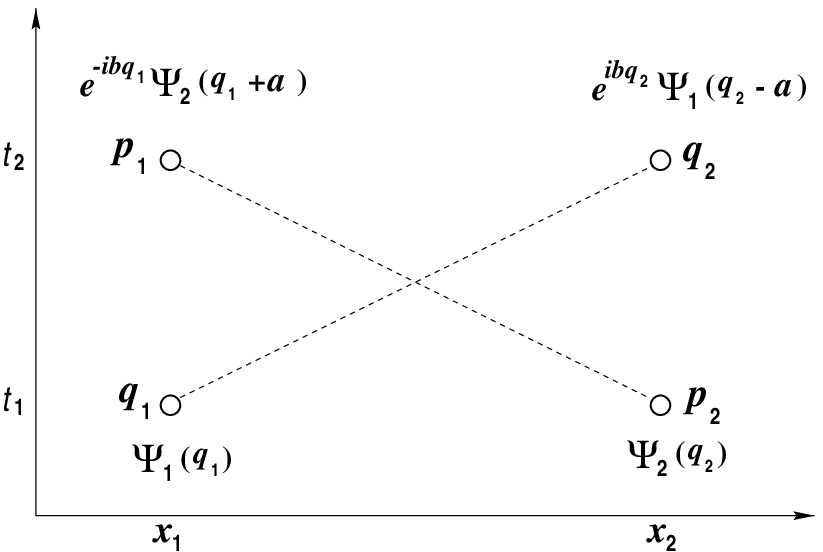} \end{center}

\noindent 
{\small {\bf \bf Fig. 3. Space-time diagram of ``crossed'' nonlocal
  measurements which result in two-way teleportation of quantum states
  of quantum systems with continuous variables.}~  Space-time locations of
  local couplings are shown.  When the nonlocal measurements
  (\ref{cross-conti}) are completed, the states of the two particles
  are interchanged up to the known shifts in $q$ and $p$.  }

\noindent
The state of particle 2 after $t_2$ is the initial state of the
particle 1 shifted by $-a$ in $q$ and by $-b$ in $p$.  Similarly, the
state of particle 1 is the initial state of particle 2 shifted by $a$
in $q$ and by $b$ in $p$.  After transmitting the results of the local
measurements, $a$ and $b$, the shifts can be corrected (even if the
quantum state is unknown) by appropriate kicks and back shifts, thus
completing a reliable teleportation of the state $\Psi_1(q_1)$ to
$\Psi_1(q_2)$ and of the state $\Psi_2(q_2)$ to $\Psi_2(q_1)$.

Surprisingly, the implementation of the reliable teleportation of
continuous variables is possible. Braunstein and Kimble made
a realistic proposal for teleporting the quantum state of a single mode of
the electro-magnetic field \cite{BK}. 
This remarkable result is an implementation of a variation of the scheme
described above which achieves a  one-way teleportation.  In their method $q$
is ``$x$''defined for a single mode of an electro-magnetic field, and
correspondingly $p$ is the conjugate momentum of $x$. The analog of
the EPR pair is obtained by shining squeezed light with a certain $x$
from one side and squeezed light with a certain $p$ from the other side
onto a simple beam splitter. The analog of the local Bell measurement is
achieved using another beam splitter and homodyne detectors. The
shifts in $x$ and $p$ which complete the teleportation procedure can
be done by combining the output field with the coherent state of 
appropriate amplitude fixed by the results of the homodyne
measurements.
  Note also a related
proposal \cite{Mol1,Mol2} for teleporting a single-photon wave packet.

Very recently the Braunstein-Kimble proposal for implementation of
continues variable teleportation \cite{tele-V} has been performed in
California Institute of Technology \cite{Furu}.  This is the first
reliable teleportation experiment. The meaning of ``reliable''
(``unconditional'' in \cite{Furu}) is that theoretically it is always
successful. It is the first experiment in which the final stage of
teleportation, i.e., transmission of the classical information to Bob
and the appropriate transformation which results in the appearance of the
teleported state  in the Bob's site, has
been implemented.  The
weakness of this experiment is that the teleported state is 
significantly distorted. The main reason for low fidelity is the
degree of squeezing of the  light which controls the quality
of the EPR pairs, the quantum channel of the teleportation.
Significant improvement of the squeezing parameters is a very difficult
technological problem. Thus, in this type of experiment one cannot
reach the high fidelity of (conditional) teleportation experiments of
photon polarization states.

One may note an apparent contradiction between the proof of Section
\ref{proof} that 100\% efficient teleportation cannot be achieved
using linear elements and single-particle state detectors and the
successful {\em reliable} teleportation experiment of the state of the
electro-magnetic field which involved only beam-splitters and local
measuring devices reported above.  Indeed, it is natural to assume
that if reliable teleportation of a quantum state of a two-level
system is impossible under certain circumstances, it should certainly
be impossible for quantum states of systems with continuous variables.
However, although it is not immediately obvious, the circumstances are
very different.  There are numerous differences.  The analog of the
Bell operator for continuous variables does not have among its
eigenvalues four states of the general form (\ref{Bell}) where
$|{\uparrow}\rangle$ and $|{\downarrow}\rangle$ signify some
orthogonal states. Another problem is that one cannot identify ``the
particles'': In the beam-splitter {\em one} input port goes to {\em
  two} output ports. One can see a ``quantum-quantum'' interaction:
the variable $x$ of one of the output ports of the beam splitter
becomes equal to $(1/ \sqrt 2)(x_1 + x_2)$, essentially, the sum of
the quantum variables of the input ports. The absence of such
``quantum-quantum'' interactions is an essential ingredient in the
proof of Section \ref{proof}. If, however, we consider the
``particles'' to be photons (which do not interact with one another)
then the homodyne detectors which measure $x$ are not single-particle
detectors---another constraint used in the proof.  Note also that the
Braunstein-Kimble method is not applicable directly for teleporting
$\Psi (x)$ where $x$ is a spatial position of a quantum system. An
additional quantum-quantum interaction which converts the continuous
variable of position of a particle to the variable $x$ of the
electro-magnetic mode is required.

 \section{Is There a Paradox with Teleportation?}

My complaints about the (mis)interpretation of the word ``teleportation''
in Section II shows that I am (over)sensitive about this
issue. This is because I was thinking a lot about it, resolving for
myself a paradox \cite{psa} which I, as a believer in the Many-Worlds
Interpretation (MWI) \cite{MWI-V} had with this experiment.
 
Consider teleportation, say in the BBCJPW scheme. We perform some
action in one place and the state is immediately teleported, up  a  local transformation (``rotation''), to an
arbitrary distant location.  But relativity theory teaches us that
anything which is physically significant cannot move faster than
light.  Thus it seems that it is the classical information (which
cannot be transmitted with superluminal velocity) about the kind of
back ``rotation'' to be performed for completing the teleportation
which is the only essential part of the quantum state.  However, the
amount of the required classical information is very small. Is the
essence of a state of a spin-1/2 particle just 2 bits?

I tend to attach a lot of physical meaning to a quantum state.  For
me, a proponent of the MWI, everything is a quantum state. But I also
believe in relativistic invariance, so only entities which cannot
move faster than light have physical reality. Thus, teleportation
poses a serious problem to my attitude. I was ready to admit that
``I'' am just a quantum state of $N\sim 10^{30}$ particles. This is
still a very rich structure: a complex function on ${\cal R}^N$. But
now I am forced to believe that ``I'' am just a point in the ${\cal
  R}^{2N}$ ?!
 
The resolution which I found for myself is as follows: In the
framework of the MWI, the teleportation procedure does not move the
quantum state: the state was, in some sense, in the remote location
from the beginning.  The correlated pair, which is the necessary item
for teleportation, incorporates all possible quantum states of the
remote particle, and, in particular, the state $\Psi$ which has to be
teleported. The local measurement of the teleportation procedure
splits the world in such a manner that in each of the worlds the state
of the remote particle differs form the state $\Psi$ by some known
transformation.  The number of such worlds is relatively small.  This
explains why the information which has to be transmitted for
teleportation of a quantum state---the information which world we need
to split into, i.e., what transformation has to be applied---is much
smaller than the information which is needed for the creation of such a
state.  For example, for the case of a spin-1/2 particle there are
only 4 different worlds, so in order to teleport the state we have to
transmit just 2 bits.  As for teleporting myself, the number of
worlds is the number of distinguishable (using measuring
devices and our senses) values of $x_i$ and $p_i$ for all continues
degrees of freedom of my body.
 
Teleportation of people will remain a dream for the foreseeable future.
First, we have to achieve the reliable teleportation of an unknown
quantum state of an external system with reasonable fidelity which is
also only a dream today.  Although the teleportation of an unknown
quantum state has not yet been achieved, the current experiments
clearly demonstrate that it can be done.  I urge the experimenters to
perform a persuasive teleportation experiment: Carol gives to Alice
(single) particles in different states (unknown to Alice), Alice
teleports the states to Bob, Bob gives them back to Carol who tests
that what she gets is what she has sent before.

I am grateful for very useful correspondence with Chris Adami, Gilles
Brassard, Samuel Braunstein, John Calsamiglia, Lior Goldenberg, Daniel
Lidar, Sergey Molotkov, Harald Weinfurter, Asher Peres, Sandu Popescu,
and Anton Zeilinger. The research was supported in part by grant
471/98 of the Basic Research Foundation (administered by the Israel
Academy of Sciences and Humanities).

\end{document}